\documentclass[10pt,letterpaper]{article}
\usepackage[latin9]{inputenc}
\usepackage{units}
\usepackage{amsmath}
\usepackage{amssymb}
\usepackage{graphicx}

\makeatletter

\usepackage{osameet2}

\newcommand{\lab}[3]{\begin{picture}(0,0)(0,0)\put(#1,#2){#3}\end{picture}}

\usepackage{cite}

\usepackage[printonlyused]{acronym}
\acrodef{MAP}{maximum a posteriori probability}
\acrodef{QAM}{quadrature amplitude modulation}
\acrodef{QPSK}{quadrature phase-shift keying}
\acrodef{ISI}{intersymbol interference}
\acrodef{GLME}{Gelfand-Levitan-Marchenko equation}
\acrodef{NFDM}{nonlinear frequency-division multiplexing}
\acrodef{NFT}{nonlinear Fourier transform}
\acrodef{FNFT}{forward NFT}
\acrodef{BNFT}{backward NFT}
\acrodef{BDF-NFDM}{BNFT based decision feedback NFDM}
\acrodef{OFDM}{orthogonal frequency-division multiplexing}
\acrodef{TX}{transmitter}
\acrodef{RX}{receiver}
\acrodef{FT}{Fourier transform}
\acrodef{DAC}{digital-to-analog converter}
\acrodef{ADC}{analog-to-digital converter}
\acrodef{LP}{Layer-Peeling}
\acrodef{SNR}{signal-to-noise ratio}
\acrodef{NIS}{nonlinear inverse synthesis}
\acrodef{DBP}{digital backpropagation}
\acrodef{QAM}{quadrature amplitude modulation}
\acrodef{SNR}{signal to noise ratio}
\acrodef{SER}{symbol error rate}
\acrodef{SE}{spectral efficiency}
\acrodef{EDC}{electronic dispersion compensation}
\acrodef{NLSE}{nonlinear Schr�dinger equation}
\acrodef{AWGN}{additive white Gaussian noise}
\acrodef{GVD}{group velocity dispersion}

\makeatother

\begin{document}

\title{A Novel Detection Strategy for Nonlinear Frequency-Division Multiplexing }

\author{Stella Civelli$^{1,2*}$, Enrico Forestieri$^{1,2}$, Marco Secondini$^{1,2}$}

\address{1 TeCIP Institute, Scuola Superiore Sant'Anna, via G. Moruzzi 1,
56124 Pisa, Italy\\
2 Photonic Networks and Technologies Nat'l Lab, CNIT, Pisa, Italy}

\email{$*$stella.civelli@santannapisa.it}
\maketitle
\begin{abstract}
A novel decision feedback detection strategy exploiting a causality
property of the nonlinear Fourier transform is introduced. The novel
strategy achieves a considerable performance improvement compared
to previously adopted strategies in terms of Q-factor.
\end{abstract}

\vspace{6pt}


\section{Introduction}

In recent years, novel transmission schemes based on the \ac{NFT}
\cite{ablowitz1981} are attracting attention as a way to tame optical
fiber nonlinearity, which limits the capacity of current optical fiber
transmission systems. Indeed, \ac{NFDM} uses the \ac{NFT} to encode
information on the nonlinear spectrum, that evolves in a very simple
way along the optical channel, free from any deterministic dispersive
and nonlinear interference. However, due to their novelty and complexity,
\ac{NFT}-based systems are affected by some drawbacks still to be
addressed. For a complete review about \ac{NFT} and \ac{NFT}-based
transmission schemes we refer to \cite{turitsyn2017optica}.

A specific and promising implementation of \ac{NFDM} is the \ac{NIS}
technique based on vanishing boundary conditions and modulation of
the continuous spectrum \cite{le2014nonlinear}. \ac{NIS} is a nonlinear
analogue of OFDM, from which it is obtained by replacing the inverse
and direct FFT operations with a \ac{BNFT} and \ac{FNFT}, respectively.
It can be easily combined with spectrally efficient modulation formats,
such as \ac{QAM}, and  with a modulation of the discrete spectrum
to further increase spectral efficiency \cite{aref2016demonstration}.
A major issue concerning \ac{NIS} is the need of operating in burst
mode by inserting a certain number of guard symbols between bursts
of information symbols to emulate the vanishing boundary conditions
of the \ac{NFT} theory. This is an issue because guard symbols reduce
the overall spectral efficiency. While the minimum number of guard
symbols is set by the memory of the channel (in particular, by the
accumulated dispersion), the number of information symbols can, in
principle, be increased at will to mitigate the efficiency loss. However,
it has been observed that the performance of \ac{NIS} systems decreases
as the burst length increases, such that only short burst lengths
can be practically considered, with a significant reduction of spectral
efficiency \cite{civelli2017noise}. This peculiar effect, not present
in conventional systems, is caused by a sort of signal-noise interaction
taking place at the \ac{RX} when computing the \ac{NFT} of the received
noisy signal \cite{Turitsyn_nature16}. The effect is so detrimental
to mask all the potential advantages of \ac{NFDM} in terms of robustness
to nonlinear interference. As a consequence, \ac{NFDM} can not yet
be considered an attractive replacement to conventional systems.

In this work, we propose a novel detection strategy for \ac{NFDM}
that, by exploiting a decision feedback scheme based on the \ac{BNFT}
(DF-BNFT), avoids the detrimental signal-noise interaction at the
\ac{RX}. Through numerical simulations, we show that the considered
detection strategy achieves a Q-factor improvement of more than $7$\,dB
compared to the previously proposed strategy based on the \ac{FNFT}.

\section{System description}

\begin{figure}
\begin{centering}
\raisebox{3mm}{\includegraphics[width=0.48\columnwidth]{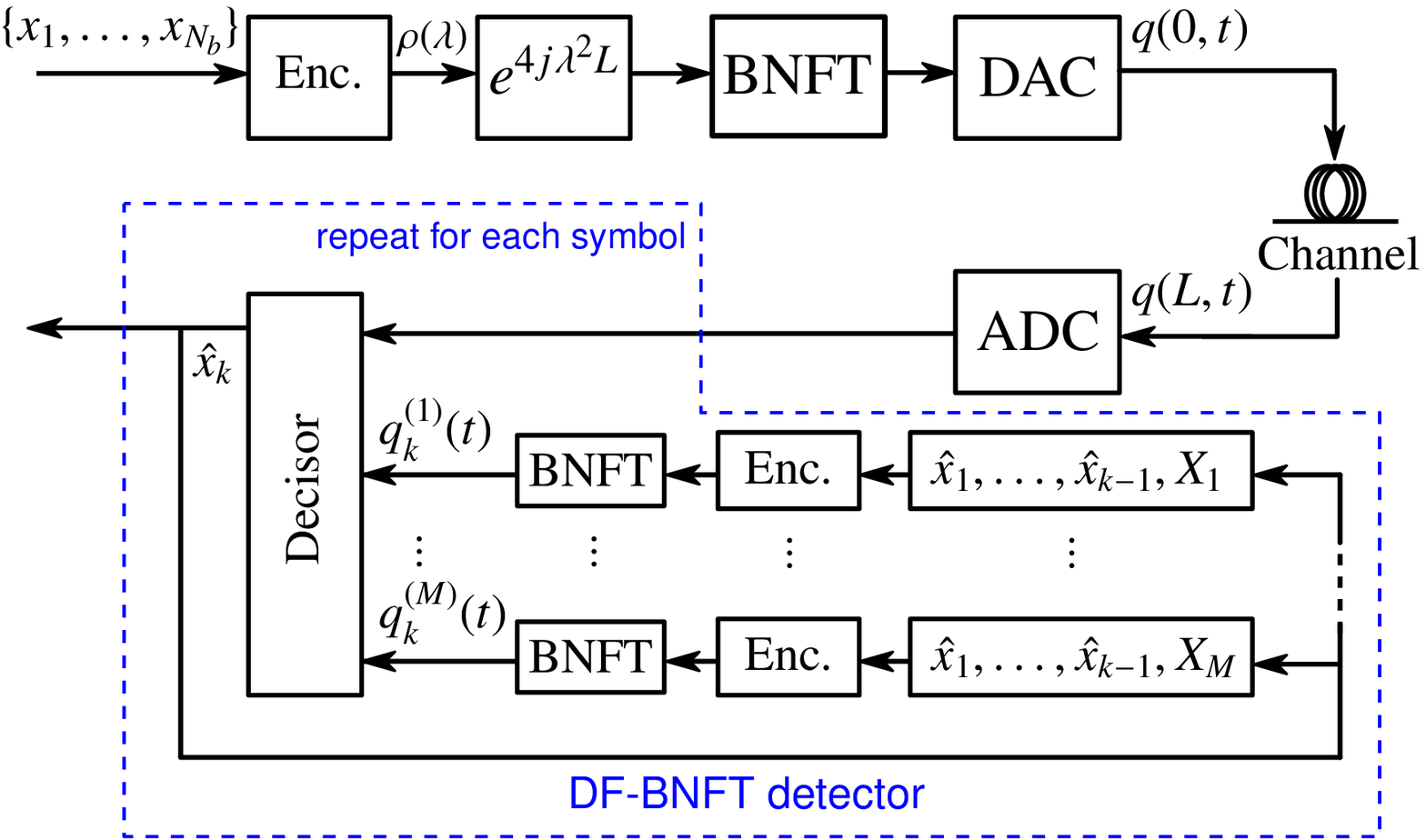}}\quad{}\includegraphics[width=0.48\columnwidth]{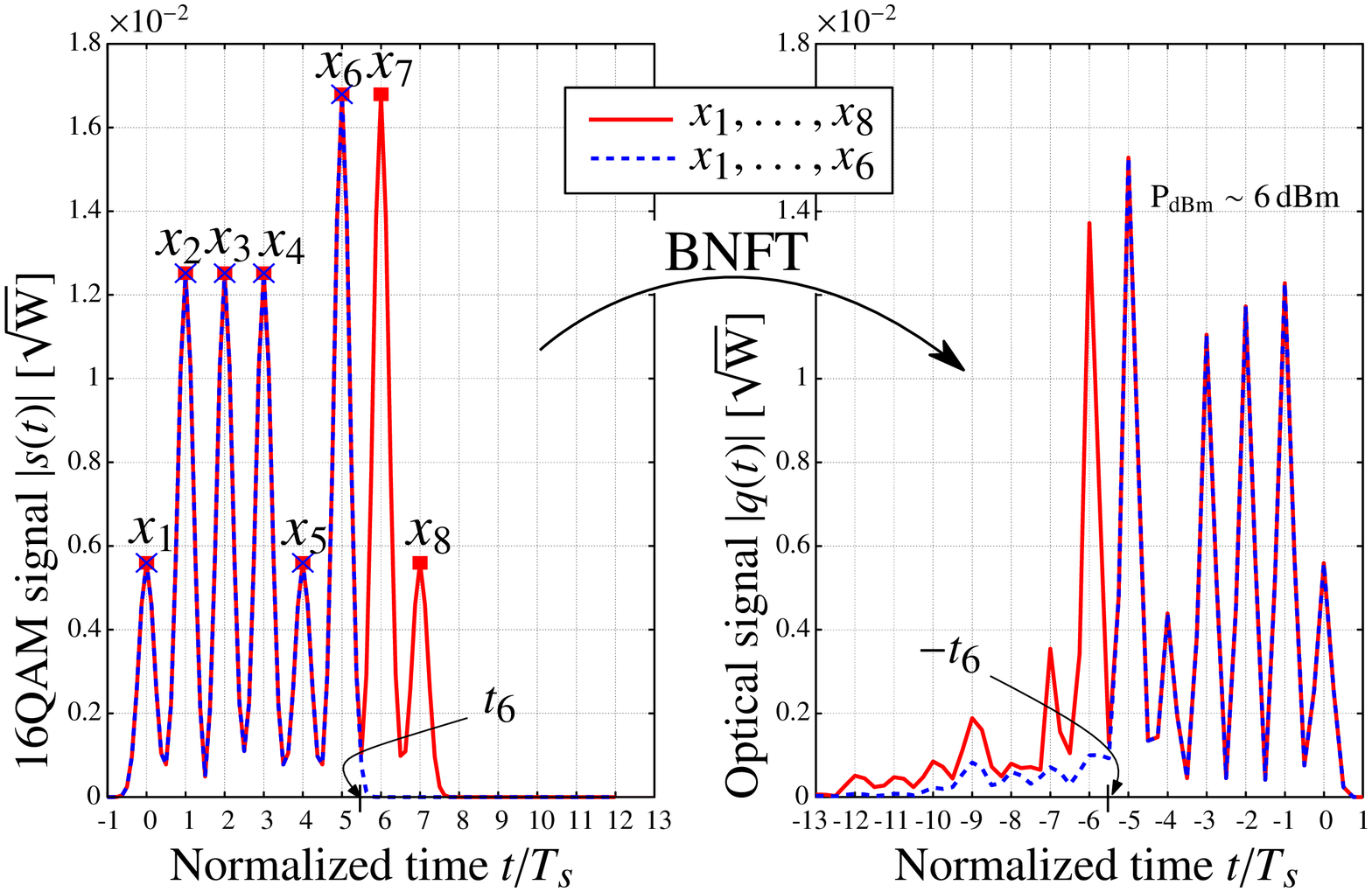}
\par\end{centering}
\begin{centering}
\hfill{}(a)\hfill{}\hfill{}(b)\hfill{}
\par\end{centering}
\caption{(a) \label{fig:FBNFDM}\protect\ac{NFDM} system with the novel DF-BNFT
detection strategy; (b) \label{fig:GLMEproperty}example of the \protect\ac{NFT}
causality property. A train of Gaussian pulses, modulated with 16\protect\ac{QAM}
symbols and almost \protect\ac{ISI}-free, before (on the left) and
after (on the right) the \protect\ac{BNFT} is shown. The solid red
and the dashed blue curves refer to the case in which only $8$ or
$6$ symbols, respectively, are considered to generate the signal.}
\vspace*{-2ex}
\end{figure}

The transmission scheme considered in this work is sketched in Fig.~\ref{fig:FBNFDM}(a).
As in the \ac{NIS} scheme \cite{le2014nonlinear}, the \ac{TX} encodes
a burst of $N$ symbols~$\{x_{1},\ldots,x_{N_{b}}\}$ drawn from
the $M$-ary \ac{QAM} alphabet $\{X_{1},\ldots,X_{M}\}$ onto a QAM
signal $s(t)$, whose ordinary Fourier transform is then mapped on
the continuous part of the nonlinear spectrum $\rho(\lambda)$. Furthermore,
before computing the \ac{BNFT} to obtain the samples of the corresponding
optical signal $q(0,t)$, deterministic propagation effects (dispersion
and nonlinearity) are precompensated by multiplying the nonlinear
spectrum by $\exp(j4\lambda^{2}L)$, where $L$ is the link length. 

In conventional NIS, the \ac{RX} recovers a noisy version of the
transmitted nonlinear spectrum $\rho(\lambda)$ by computing the \ac{FNFT}
of the received optical signal $q(L,t)$, and then makes decisions
based on standard matched filtering and symbol-by-symbol detection.
The improved detection scheme proposed in this work originates from
the idea that, since a detrimental signal-noise interaction takes
place when computing the \ac{FNFT} of the received noisy signal,
decisions could be alternatively made by comparing the received signal
with the \ac{BNFT} of all the possible transmitted (noiseless) waveforms,
avoiding signal-noise interaction. Selecting the waveform (and the
corresponding symbols) closest to the received optical signal would
correspond to a \ac{MAP} strategy, under the assumption that the
accumulated optical noise can be modelled as \ac{AWGN}. To avoid
an exponential growth of the detector complexity with the burst length
$N_{b}$, a causality property of the \ac{NFT} (more on this later)
and a decision feedback scheme are finally employed, obtaining the
DF-BNFT detection scheme depicted in Fig.~\ref{fig:FBNFDM}(a). The
$N_{b}$ symbols of each burst are iteratively detected according
to\setlength{\abovedisplayskip}{2pt}\setlength{\belowdisplayskip}{1pt}
\[
\hat{x}_{k}\mathrm{=\underset{{\scriptstyle X_{i}\in\{X_{1},..,X_{M}\}}}{argmin}}\int_{-t_{k}}^{-t_{k-1}}\!|q(t)-q_{k}^{(i)}(t)|^{2}\,\mathrm{d}t,\quad\quad\mathrm{for}\quad k=1,\ldots,N_{b}
\]
where $q_{k}^{(i)}(t)$ is the trial waveform obtained by applying
the same encoding and \ac{BNFT} operations of the \ac{TX} to the
sequence $\hat{x}_{1},..,\hat{x}_{k-1},X_{i}$; $\hat{x}_{1},..,\hat{x}_{k-1}$
are the previously detected symbols; $t_{k}=(k-1/2)T_{s}$; and $T_{s}$
is the symbol time.

The causality property of the \ac{NFT}, employed to derive the above
detection strategy, can be derived from the Gelfand-Levitan-Marchenko
equation (an integral equation used to compute the \ac{BNFT} \cite{ablowitz1981})
and relates the optical signal $q(t)$ to the inverse Fourier transform
of its nonlinear spectrum $s(t)$. The property ensures that $q(t)$
for $t>-\tau$ depends only on $s(t)$ for $t<\tau$. Thus, if $s(t)$
is \ac{ISI}-free, the optical signal received after $-t_{k}$ depends
only on the first $k$ symbols $x_{1},..,x_{k}$. Therefore, one can
take a decision on the $k$-th symbol by using only the portion of
signal received after $-t_{k}$ and the previously decided symbols
$\{\hat{x}_{1},\ldots,\hat{x}_{k-1}\}$, avoiding \ac{ISI}. Unfortunately,
as the converse property does not hold\textemdash the signal received
before $-t_{k}$ does not depend only on the remaining symbols $x_{k+1},..,x_{N_{b}}$\textemdash the
proposed strategy is suboptimum. The application of the property to
our scheme is sketched in Fig.~\ref{fig:GLMEproperty}(b), where
it is shown that the optical signal generated from a sequence of 8
16-QAM symbols, and the signal generated by only the first 6 symbols
of the same sequence, are equal after time $-t_{6}$. However, the
shorter optical signal has a non-zero tail before $-t_{6}$ even if
the corresponding QAM signal vanishes after $t_{6}$. As regards the
computational cost, the detection of each burst requires $M$ \ac{BNFT}
of the whole signal (at each step, trial waveforms are computed only
in a short time window), while the conventional FNFT-based strategy
uses only $1$ \ac{FNFT}.

\section{System performance}

System performance has been evaluated through simulations for a 16-QAM
modulation format. A Gaussian pulse shape with normalized root mean
square width $0.2$ is chosen, such that the resulting QAM signal
$s(t)$ is practically \ac{ISI}-free. The symbol rate is $R_{s}=1/T_{s}=\unit[50]{GBd}$.
The fiber channel, whose length is $L=\unit[2000]{km}$, is characterized
by group velocity dispersion parameter $\beta_{2}=\unit[-20.39]{ps^{2}/km}$,
attenuation $\alpha=\unit[0.2]{dB/km}$, and nonlinear coefficient
$\gamma=\unit[1.22]{W^{-1}km^{-1}}$. Polarization effects are neglected
and ideal distributed amplification with spontaneous emission factor
$\eta_{\mathrm{sp}}=4$ is considered along the channel. This idealized
scenario ensures that the investigated effect is not masked by other
propagation effects, and that its impact and mitigation can be clearly
observed. The bandwidth of both the \ac{DAC} and the \ac{ADC} is
$\unit[100]{GHz}$. To account for dispersion, $2000$ guard symbols
separate different bursts. As customary, the performance is expressed
in terms of the Q-factor, defined as $Q_{\mathrm{dB}}^{2}=20\log_{10}[\sqrt{2}\mathrm{erfc}^{-1}(2P_{b})]$,
where $P_{b}$ is the bit error probability measured by direct error
counting. All digital operations, in particular both the \ac{BNFT}
and the \ac{FNFT}, were computed with the accuracy required to avoid
any numerical impact on simulation results. The rate efficiency term
$\eta=N_{b}/(2000+N_{b})$ was used to take into account the spectral
efficiency loss due to the insertion of $2000$ guard symbols, as
previously explained.

\begin{figure}
\begin{centering}
\lab{110}{164}{(a)}\lab{350}{164}{(b)}\includegraphics[width=0.48\columnwidth]{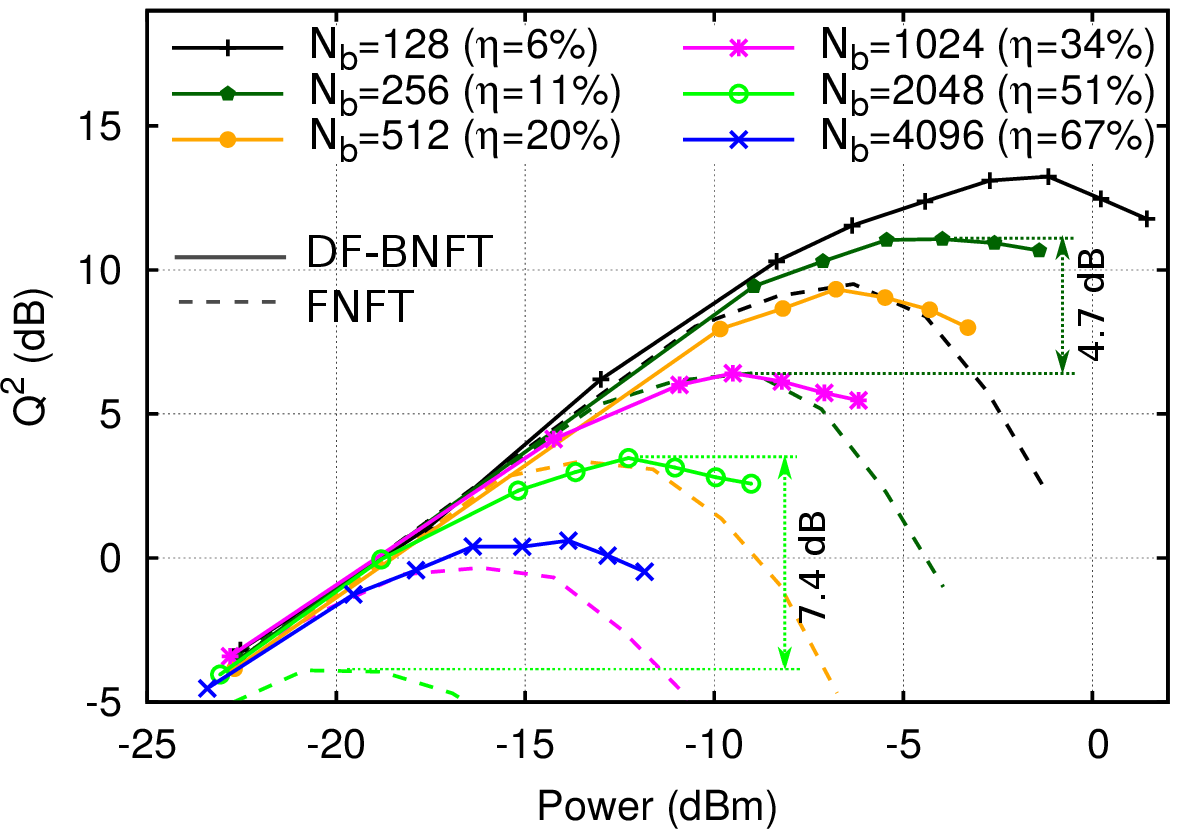}\quad{}\includegraphics[width=0.48\columnwidth]{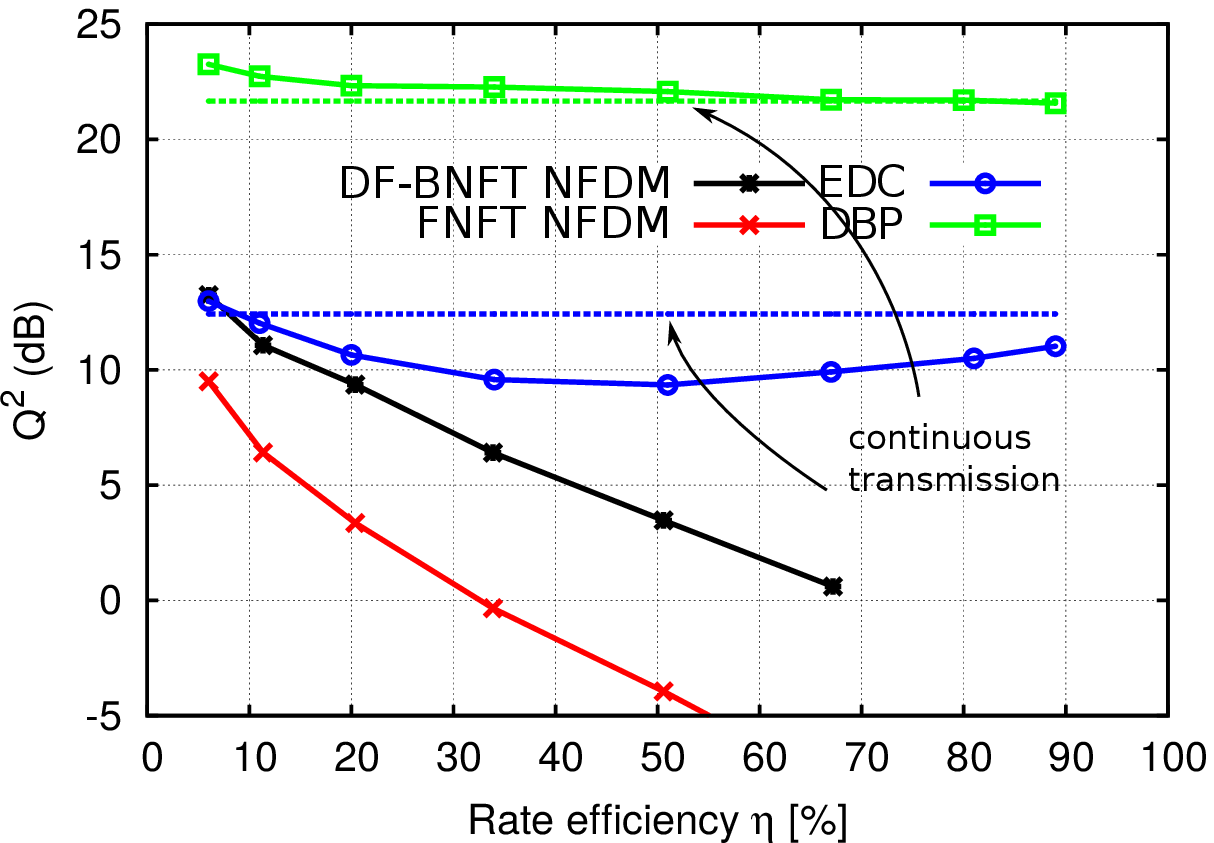}\vspace*{-1ex}
\par\end{centering}
\caption{(a) \label{fig:FBNFDMperf}System performance for different burst
length as a function of the optical power, obtained with DF-BNFT and
\protect\ac{FNFT} detection (same color same length); (b) \label{fig:FBNFDM_conf}Optimal
performance of DF-BNFT and standard \protect\ac{FNFT} detection for
\protect\ac{NFDM}, and conventional systems as a function of the
rate efficiency $\eta$. }
\vspace*{-3ex}
\end{figure}

Fig.~\ref{fig:FBNFDMperf}(a) reports the system performance as a
function of the optical power, for the FNFT (dashed lines) and proposed
DF-BNFT (solid lines) detection and different burst lengths. Firstly,
Fig.~\ref{fig:FBNFDMperf}(a) shows the typical behavior of \ac{NIS},
confirmed also by theoretical studies \cite{Turitsyn_nature16}: the
higher the burst length, i.e., the rate efficiency, the worse the
performance. Secondly, the figure highlights a significant performance
improvement of DF-BNFT with respect to FNFT detection ($4.7$dB for
$\eta=11\%$, $7.4$dB for $\eta=51\%$). Thirdly, despite the improvements,
performance decay remains. 

Fig.~\ref{fig:FBNFDMperf}(b) compares, as a function of the rate
efficiency, the best performance (at optimum power) obtained by NFDM
systems using the two considered detection schemes with those obtained
by conventional systems using ideal \ac{EDC} and \ac{DBP}. As can
be seen, the improvement achieved by DF-BNFT detection, though significant
with respect to FNFT detection, is still not sufficient to outperform
conventional systems, as performance decay continues to worsen, rather
than saturate as in conventional systems.

\section{Conclusions}

This work introduces a novel detection strategy for \ac{NFDM} based
on the backward \ac{NFT} and a decision feedback scheme. The proposed
DF-BNFT technique allows for a considerable advantage (more than 7\,dB)
with respect to a standard detection based on the forward \ac{NFT}.
The improvement, though not yet sufficient to make \ac{NFDM} competitive
with conventional systems, demonstrates that the critical NFDM limitations
due to signal-noise interaction at the \ac{RX} can be overcome, and
paves the way for the advent of transmission paradigms customized
for the nonlinear optical channel. Further improvements and complexity
reduction of the proposed technique are currently under investigation.

\bibliographystyle{osajnl}
\bibliography{ref}

\end{document}